\documentclass[12pt]{iopart}
\usepackage[dvips]{epsfig}
\begin{document}

\title[]{Dipolar oscillations in a quantum degenerate Fermi-Bose atomic mixture}

\author{F. Ferlaino, R. J. Brecha \dag\, P. Hannaford \ddag\, F. Riboli, G. Roati, G. Modugno, M. Inguscio
 }

\address{European Laboratory for Nonlinear Spectroscopy, Universit\'a di Firenze,
and Istituto Nazionale per la Fisica della Materia, Via Nello
Carrara 1, 50019 Sesto Fiorentino, Italy.\\
\dag\ Physics Department, University of Dayton, Dayton, OH
45469-2314 USA.\\
\ddag\ Swinburne University of Technology, Mail H31, PO Box 218,
Melbourne, Australia.}

\begin{abstract}
We study the dynamics of coupled dipolar oscillations in a
Fermi-Bose mixture of $^{40}$K  and $^{87}$Rb atoms. This
low-energy collective mode is strongly affected by the
interspecies interactions. Measurements are performed in the
dynamical and quantum degenerate regimes and evidence the crucial
role of the statistical properties of the mixture. At the onset of
quantum degeneracy, we investigate the role of Pauli blocking and
superfluidity for K and Rb atoms, respectively, resulting in a
change in the collisional interactions.
\end{abstract}


\submitto{\JOB}

\maketitle

\section{Introduction}
Recent research on ultracold atomic mixtures has opened a new
scenario in the field of quantum degenerate gases. Thanks to
sympathetic cooling techniques, it is now possible to study a
variety of degenerate mixtures composed of two bosonic species
\cite{science,bb} as well of bosonic and fermionic isotopes of the
same species \cite{hulet,salomon} or of different species
\cite{had,fb}. In a mixture, the interactions between unlike
particles determine relevant properties such as the efficiency of
rethermalization processes or the stability of the mixture against
collapse \cite{molmer,collapse}. They also could play a crucial
role in the study of atoms-to-molecule association \cite{wieman}
and in the binding energy for BCS pairing of fermions
\cite{viverit}. The elastic interactions between ultracold gases
can be investigated by the study of the damping of dipolar
oscillations. This method was originally used in \cite{maddaloni}
for a two components Bose-Einstein condensate (BEC) loaded in
different hyperfine levels and was then extended to a gas of
fermions in two spin-states \cite{gens} and to a mixture composed
of different atoms \cite{coll}. In a harmonic potential, single
gases undergo undamped collective oscillations, whereas two gases
experiencing different confinements can exhibit a damped
out-of-phase motion giving quantitative information about the
scattering processes. In the degenerate regime, the dipolar
oscillations are affected by the quantum statistic and their study
can give useful information on the mean-field interactions between
the two gases.

In the present work we apply for the first time this technique to
a degenerate Fermi-Bose mixture. The study of $^{40}$K-$^{87}$Rb
dipolar oscillations gives a signature of the strong interspecies
interactions. We investigate how the damping is modified when one
or both gases are brought to the quantum degenerate regime. We
observe a transition from hydrodynamic to collisionless regime
which reveals the roles of Pauli blocking of collisions at the
onset of Fermi degeneracy. At lower temperatures, when also BEC
occurs, we evidence also the effect of superfluidity of bosons on
the damping.

\section{Experimental}
The ultracold K-Rb mixture is produced by a combination of
magneto-optical trapping and sympathetic cooling in a magnetic
trap. The experimental apparatus and techniques are described in
detail in Refs.~\cite{science,fb}. In brief, about 10$^{9}$ Rb
atoms and 10$^5$ K atoms at a temperature $T$=100~$\mu$K are
loaded in a quadrupole-Ioffe configuration (QUIC) magneto-static
trap. Here Rb-selective evaporative cooling is performed using
radio-frequency radiation, taking advantage of the different
Zeeman structure of the two species. Potassium atoms are
sympathetically cooled through elastic interspecies collisions.
With an evaporation ramp lasting about 25~s we are able to cool
typically a few 10$^{4}$  atoms to the quantum degenerate regime.
For these numbers of atoms, the Fermi temperature for K is of the
order of $T_F$=300~nK and the critical temperature for BEC of Rb
is $T_c$=150~nK. In these conditions, the BEC is usually
completely surrounded by the Fermi gas, whose dimensions are
approximatively as twice as large \cite{fb}. Both species are
trapped in their stretched spin states, $|F=9/2, m_F=9/2\rangle$
for K and $|2, 2\rangle$ for Rb. These states experience the same
trapping potential, with axial and radial harmonic frequencies
$\omega_{a}=2\pi \times 24$~s$^{-1}$ and $\omega_{r}=2\pi\times
317$~s$^{-1}$ for K, while those for Rb are a factor
$(M_{Rb}/M_K)^{1/2}\approx 1.47$ smaller. The different trapping
frequencies experienced by the two gases allows one to induce a
relative motion between the two components.

To excite dipolar oscillations of the two samples we suddenly
displace the minimum of the trapping potential in the horizontal
direction $x$, by changing the ratio of currents in the trap
coils. By an appropriate choice of the amplitude and timing of
such displacement we can excite a quasi-pure dipolar oscillation,
with no apparent higher-order (shape) oscillations. The typical
mean relative velocity of K and Rb samples during the subsequent
oscillations is $\sqrt{\langle v^2 \rangle}$=5~$\mu$m/ms. We note
 that the amplitude of oscillations is small enough to preserve
the overlap of the two clouds even in the degenerate regime
\cite{fb}.

The centers-of-mass positions of both samples in the magnetic trap
can be determined simultaneously at the end of each experimental
run by means of two-color absorption imaging. For this, we use two
delayed light pulses, at 767~nm for K and 780~nm for Rb, lasting
30~$\mu$s each, which are imaged on different areas of a CCD
sensor. The atoms are imaged after a ballistic expansion following
release from the trap. The expansion times for K and Rb are 4~ms
and 19~ms respectively.

\subsection{Thermal gases: K-Rb scattering length}
\label{therm}

\begin{figure}[tb]
\begin{center} \epsfxsize=14cm \epsfbox{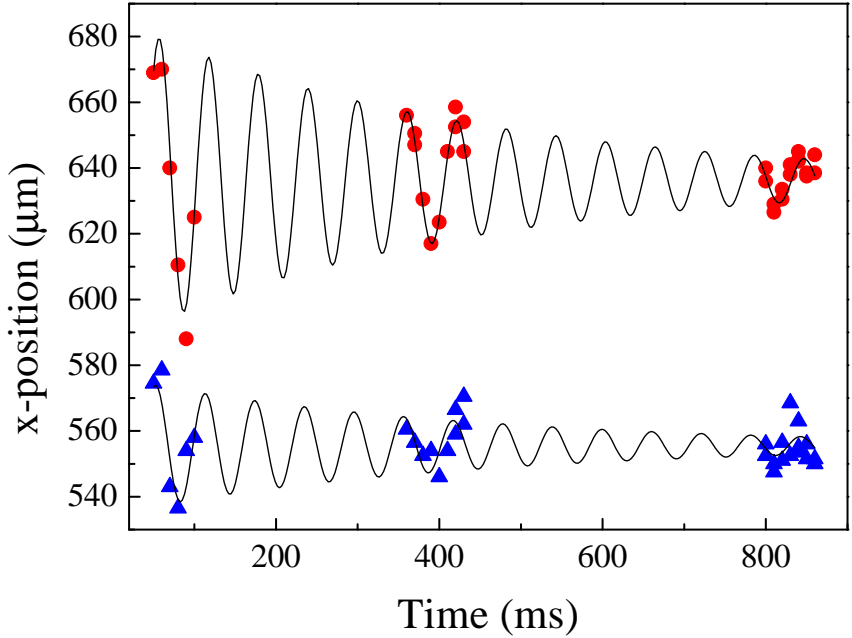}
\end{center}
\caption{Coupled dipolar oscillations of 8$\times$10$^{3}$
nondegenerate K (triangles) and 8$\times$10$^{4}$ uncondensed Rb
(circles) atoms along the horizontal axis at $T$=300~nK. The two
samples oscillate at the same frequency with a collisional rate
$\Gamma_{coll}$=190~s$^{-1}$ typical of a hydrodynamic regime. The
solid lines are the best fit to the model presented in the text. }
\label{pochi}
\end{figure}

In a first set of measurements we have studied the damping of the
dipolar oscillations for a nondegenerate K-Rb sample, at
temperatures $T$=300-500~nK. The effect of the interspecies
collisions is apparent from the data in Fig.~\ref{pochi}: both K
and Rb oscillations are damped, and the K motion is also
frequency-shifted.

To describe the coupled center-of-mass motion, we use a classical
model for two harmonic oscillators coupled through a viscous
damping. The coupled equations of motion are
\begin{eqnarray}
\ddot{x}_{Rb} &=& -\omega_{Rb}^2 x_{Rb}- \frac{4}{3} \frac{M_K}{M}
\frac{N_K}{N} \Gamma_{coll} \left( \dot{x}_{Rb} - \dot{x}_K
\right) \nonumber \\ \ddot{x}_K &=& -\omega_K^2 x_K + \frac{4}{3}
\frac{M_{Rb}}{M} \frac{N_{Rb}}{N} \Gamma_{coll} \left(
\dot{x}_{Rb} - \dot{x}_K \right)\,, \label{model}
\end{eqnarray}
where $M$ is the total mass $M_K+M_{Rb}$, $N$ the total number of
atoms $N_K+N_{Rb}$, and $\Gamma_{coll}$ is the rate of K-Rb
two-body elastic collisions. This model can be used to describe
the center-of-mass motion of the two clouds, whereas all the
microscopic damping mechanisms are described by the quantity
$\Gamma_{coll}$. Assuming two Boltzmann distributions for the
gases at sufficiently low temperature $T$ to be in the Wigner
regime, one obtains
\begin{equation}
\Gamma_{coll}={\bar n} \sigma v_{th}\,, \label{gamma}
\end{equation}
where $v_{th}=\sqrt{8 k_B T /\pi M}$ is the rms relative thermal
velocity and $\bar n = (\frac{1}{N_K}+\frac{1}{N_{Rb}})\int {n_K
n_{Rb}d^3x}$. Finally, the collision cross-section $\sigma$
depends on the interspecies s-wave scattering length as
\begin{equation}
\sigma= 4\pi a^2\,.
\end{equation}

\begin{figure}[tb]
\begin{center}
\epsfxsize=12cm \epsfbox{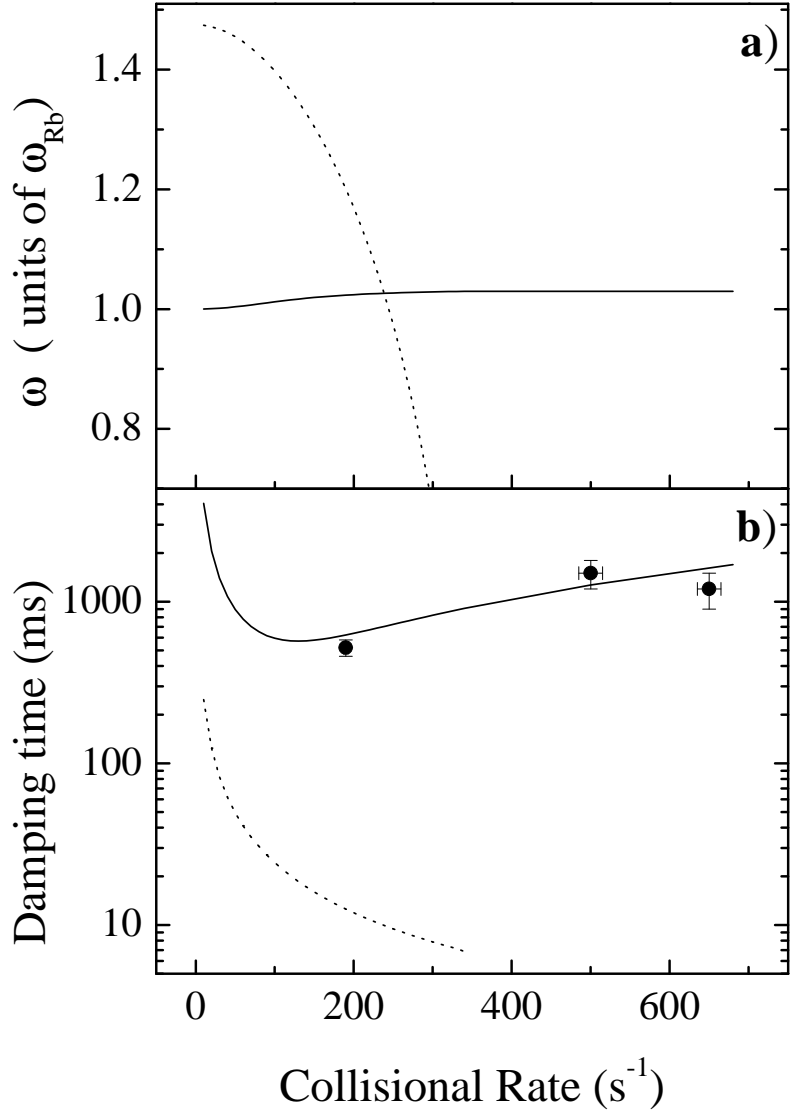}
\end{center}
\caption{Oscillation frequency (a) and damping time (b) for the
two normal modes of the K-Rb coupled dipole oscillations, as a
function of the collisional rate. The curves are calculated
according to the classical kinetic model (see Eq.~\ref{model}),
for a sample composed of $\sim$88\%  of Rb (solid line) and $\sim$
12\%  of K (dashed line), with an interspecies scattering length
$a$=-410~$a_0$. The points in (b) are the experimental data for a
thermal sample in the hydrodynamic regime. Here we are not able to
observe the strongly damped out-of-phase motion.} \label{theory}
\end{figure}

\begin{figure}
\begin{center}
\epsfxsize=12cm \epsfbox{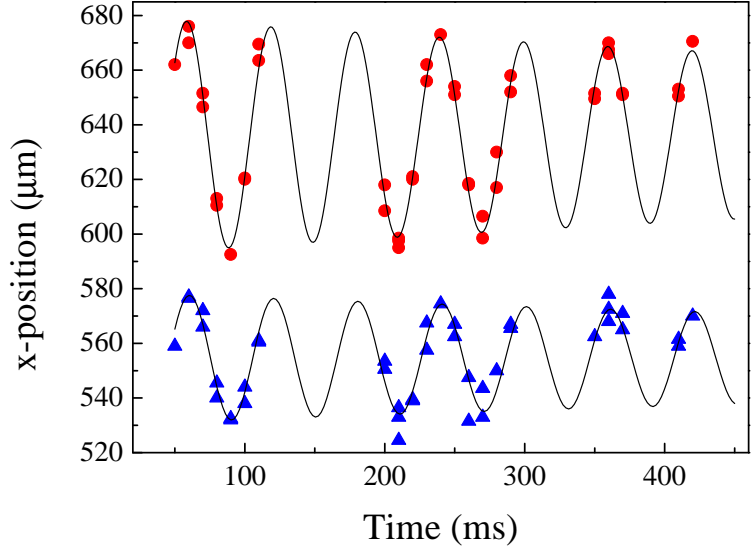}
\end{center}
\caption{\label{molti} Coupled dipolar oscillations of
2$\times$10$^{4}$ thermal  K (triangles) and 2$\times$10$^{5}$ Rb
(circles) atoms along the horizontal axis in the hydrodynamic
regime at $T$=300~nK. The two samples oscillate at the same
frequency with a collisional rate $\Gamma_{coll}$=650~s$^{-1}$.
The solid lines are the best fit to the model presented in the
text (see Eq.~\ref{model}).}
\end{figure}

The motion of two coupled harmonic oscillators is a well know
problem, and here we just recall the properties which are
important for the present experiment. In general such a system has
two normal modes, whose frequencies $\omega$ and damping time
$\tau$ vary with the collisional rate $\Gamma_{coll}$. The
behaviour, calculated by numerically solving Eq.~\ref{model} for
the ratio of atom numbers that we have typically in the
experiment, $r$=$N_{Rb}/N_K$=7.5, is shown in Fig.~\ref{theory}.
At low collisional rate, in the so-called collisionless regime
($\omega_{Rb}\tau, \omega_{K}\tau \ll$1), the two samples are
predicted to oscillate at their bare frequencies (
$\omega_{K}\approx 2\pi\times 24$ s$^{-1}$, $\omega_{Rb}\approx
2\pi\times 16.3$ s$^{-1}$ ), and the ratio of the two damping
times scale as the inverse ratio of the total mass of each sample.
As the collisional rate increases, the damping time of the two
normal modes decreases and their frequencies are shifted towards
an intermediate value. Here each sample oscillates at a
combination of both normal modes. Finally, at very high
collisional rate ($\omega\tau
>$1) the system enters the hydrodynamic regime. Here there is a
mode at this intermediate frequency with low damping and a second
overdamped mode whose frequency rapidly decreases with increasing
$\Gamma_{coll}$. This situation corresponds to what we observe in
the experiment ( see Fig.~\ref{pochi}), where the two samples
oscillate at the same frequency, almost in phase and with a long
damping time. In Fig. \ref{theory} we also report the measured
damping time vs the collisional rate, that confirm the prediction
of an increasing $\tau$ for increasing $\Gamma_{coll}$. For
example in Fig.~\ref{molti} we show the oscillations of a sample
composed of three times as many atoms of each species as that in
Fig.~\ref{pochi}. As predicted, the damping time is three times
longer than the one observed in Fig.~\ref{pochi}.

The dependence of $\tau$ on $\Gamma_{coll}$ in the hydrodynamic
regime is weaker than in the collisionless regime (see
Fig.~\ref{theory}b), while instead there is a strong dependence of
the relative phase of K and Rb motion on $\Gamma_{coll}$. The
latter dependence helps to provide an accurate determination of
the experimental $\Gamma_{coll}$, since the phase of the
oscillations can be determined with relatively high accuracy.

The overall behaviour of the normal modes of the coupled
oscillators can be qualitatively understood by noting that an
increasing collisional rate tends to restore local equilibrium in
the mixture. For the system, the way to produce this with the
lowest energy cost is to shift the oscillation frequency of the
sample with the smaller mass, and then to reduce the phase
difference between the two samples. These features do not change
significantly when changing the number ratio $r$. In particular,
we find that the minimum damping rate at the transition from the
collisionless to the hydrodynamic regime is always around
140~s$^{-1}$, whereas the region where the frequencies of the two
modes cross is centered around 220~s$^{-1}$, irrespective of $r$.
In any case, we note that this crossing occurs only for $r>$1; for
smaller ratios it is the "potassium mode" , i.e. the high
frequency mode, that acquires long damping times in the
hydrodynamic regime.

The experimental values of the collisional rate  can be used to
determine the K-Rb s-wave triplet scattering length. To obtain the
value that we have already reported in \cite{collapse}, we fit the
experimental data for the dipole oscillations with the solution of
Eq.~\ref{model}, and we extract a value for the collision
cross-section and hence for the scattering length $a$. We have
repeated this procedure by varying the temperature in the range
$T$=300-500~nK, the total number of atoms in the range
$N$=10$^4$-5$\times$10$^5$, and the ratio $r$ from 2.5 to 7.5. We
have then made a weighted average of the resulting values for the
scattering length and obtained $|a|$=410$^{+90}_{-80}$~$a_0$. Here
the uncertainty is dominated by a 40\% a priori uncertainty in the
number density.

\subsection{Fermi gas and thermal bosons: Pauli blocking of collisions}
\label{ferther}

We have repeated the above experiments at lower temperatures, with
one or both gases in the degenerate regime. As we will show in the
following, the study of dipolar oscillations in this regime can
reveal effects caused by the modification of the collisional
properties of a degenerate gas. The quantitative analysis of
experimental results is, however, complicated by the fact that at
finite $T$ the bosonic sample has both a condensed and an
uncondensed fraction, which are expected to interact in different
ways with fermions. To simplify the analysis we have first
investigated the interaction between the degenerate Fermi gas and
an uncondensed bosonic cloud.

In the experiment, this situation is actually easier to produce
than the opposite case. Indeed, for the same number of atoms of
each species, the Fermi temperature is higher than the critical
temperature for condensation also because of the different
confinements:
\begin{eqnarray}
k_B T_F&=&\hbar {\bar \omega_K} (6N_K)^{1/3}\,, \nonumber\\
k_B T_c&=&\hbar {\bar \omega_{Rb}} (N_{Rb}/1.26)^{1/3}\,.
\end{eqnarray}

To produce the mixed sample we intentionally force the final stage
of evaporation of Rb, in order to reduce the atom number and
prevent the onset of BEC. Using this procedure we have been able
to produce a sample composed of 2$\times$10$^4$ K atoms and 10$^4$
Rb atoms at $T$=120~nK, for which $T_F$=260~nK and $T_c$=110~nK.
In this condition, we can calculate the damping rate for two
classical samples $\Gamma_{coll}=$100~s$^{-1}$ which correspond to
a mixture in the hydrodynamic regime.

\begin{figure}[tb]
\begin{center}
\epsfxsize=12cm \epsfbox{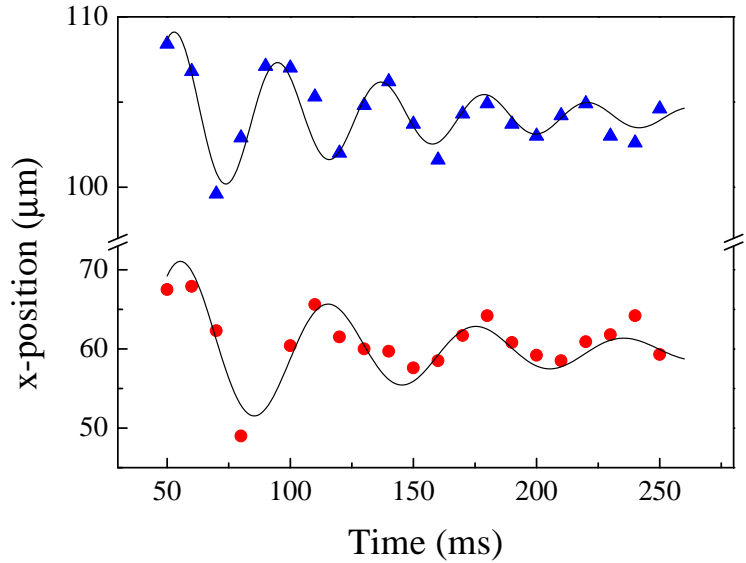}
\end{center}
\caption{Coupled dipolar oscillations of degenerate K (triangles)
and thermal Rb (circles) clouds along the horizontal axis in the
collisionless regime at $T$=120~nK. The two samples oscillate at
their bare frequencies with a collisional rate
$\Gamma_{coll}$=60~s$^{-1}$ typical of a collisionless regime. The
solid lines are the best fit to the model presented in the text.}
\label{fermitherm}
\end{figure}

In Fig.~\ref{fermitherm} we show the evolution of the coupled
dipolar oscillation for this particular sample. Note that in this
case each of the two species oscillates at its bare frequency
($\omega_{K}\approx 2\pi\times 24$~s$^{-1}$, $\omega_{Rb}\approx
2\pi\times 16.3$~s$^{-1}$); this indicates that the collisional
rate is now smaller than the oscillation frequencies, i.e. we are
in the collisionless regime. From a fit we obtain
$\Gamma_{coll}$=60(10)~s$^{-1}$, which is significantly smaller
than the one expected for a non-degenerate sample. This is a clear
indication that the interspecies collisional processes are
somewhat reduced once the Fermi gas is cooled below $T_F$. This
phenomenon is actually predicted to occur because of the Pauli
blocking of elastic collisions \cite{papp}: at $T \ll T_F$ only
the fermions in the outer shell of the Fermi sphere, having an
energy $E$=$E_F$-$E_{CM}$ can scatter with bosons and participate
in the damping. Here $E_{CM}$ is the collision energy of fermions
and bosons in the center-of-mass frame
\begin{equation}
E_{CM}=\mu \langle v^2 \rangle / 2 ,
\end{equation}
where $\mu$ is the reduced K-Rb mass. An upper value for $E_{CM}$
for the typical experimental parameters is $E_{CM}/K_B \approx$
50~nK, which is significantly smaller than the typical Fermi
energy $E_F/K_B$=300~nK. At $T$=0~K, just half of the fermions can
participate to the scattering processes with bosons, however a
finite temperature reduces Pauli blocking because of the smearing
of the Fermi distribution. As shown in \cite{ferrari}, for this
excitation energy Pauli blocking should nevertheless produce a
significant reduction of $\Gamma_{coll}$, also at the temperature
$T$=0.4$T_F$ of the experiment, as indeed we observe.

The Pauli blocking phenomenon can be viewed as a tool to
investigate the level of degeneracy of a trapped fermionic gas. In
a degenerate Fermi gas, in contrast to a thermal cloud, the
spatial density and the mean speed of the atoms remain constant
when lowering the temperature. This deviation from classical
behaviour does not allow direct measurement of the sample
temperature. Since the reduction of the damping rate is predicted
to depend on $T$, the damping of relative motion could be used to
determine the temperature of fermions even in the regime where
boson thermometry does not work anymore, as suggested in
\cite{ferrari}and in \cite{demarco}. A quantitative investigation
of the phenomenon needs to be supported by an appropriate quantum
model for the collisions.

\subsection{Fermi gas and BEC: superfluidity}
\label{ferbec}

\begin{figure}[tb]
\begin{center}
\epsfxsize=12cm \epsfbox{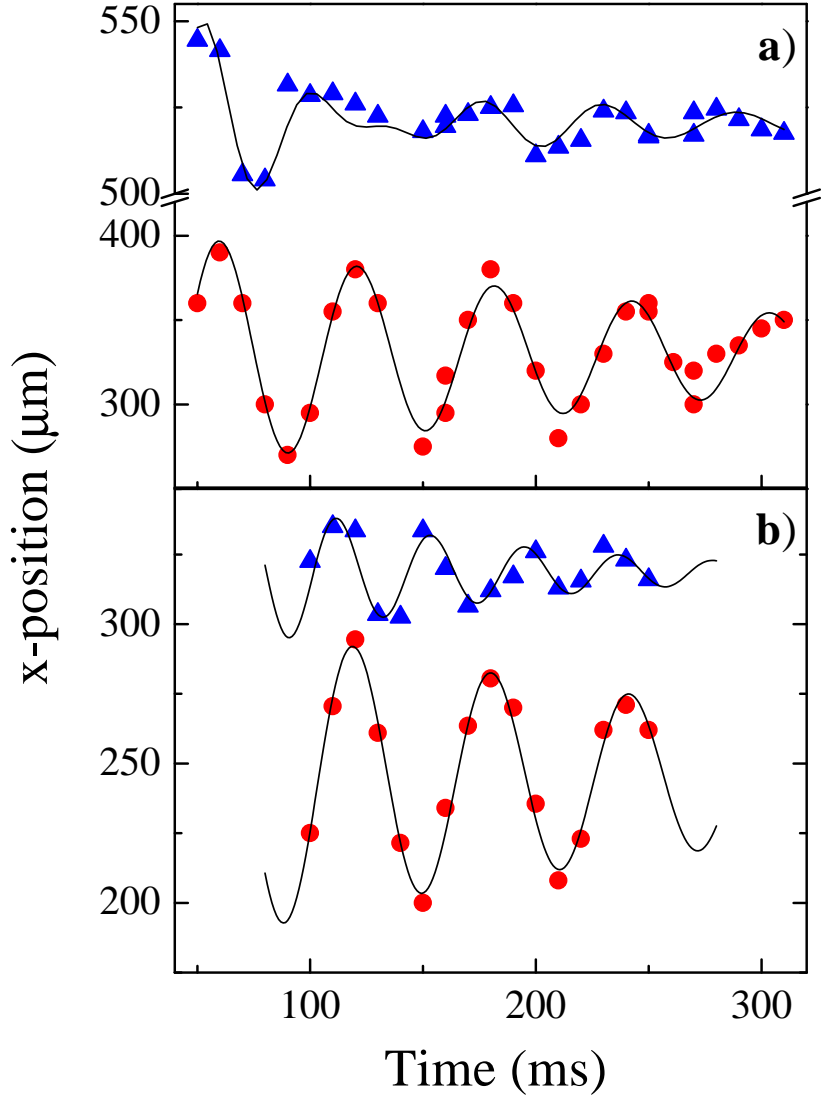}
\end{center}
\caption{\label{fbtt} Coupled dipolar oscillations of degenerate K
(triangles) and Rb BEC (circles) clouds along the horizontal axis:
(a) in the crossover between Hydrodynamic and collisionless
regimes at $T \approx$~0.84 $T_c$ with $\Gamma_{coll}$=35~s$^{-1}$
and (b) in the collisionless regime at $T <~$0.7 $T_c$ with
$\Gamma_{coll}$=50~s$^{-1}$. The solid lines are the best fit to
the model presented in the text.}
\end{figure}

Finally, we have repeated the experiments in the quantum regime
using a degenerate Fermi gas and a BEC. We observe a crossover
from the hydrodynamic to the collisionless regime when the
temperature of the bosons is lowered by evaporation below $T_c$.
In Fig.~\ref{fbtt}a, we show strongly damped Rb and K oscillations
in the crossover region between the hydrodynamic and collisionless
regimes. Here the motion of the two clouds is the superposition of
two damped oscillations at frequencies close to the bare
frequencies of the two clouds. Here the dipole oscillations are
excited for a mixed sample with $N_K$=2.5$\times 10^4$ K atoms and
$N_{Rb}$=3.9$\times 10^4$ at $T$=0.84$T_c$=118~nK. In
Fig.~\ref{fbtt}b we show a measurement performed with a BEC at
even lower a temperature $T<$0.7$T_c$. Here the two samples
oscillate at their bare frequencies, indicating that we are in the
collisionless  regime. Notice that the minimum measurable
temperature is limited to $T$=0.7$T_c$ by the minimum detectable
uncondensed Rb component. .

\begin{figure}[tb]
\begin{center}
\epsfxsize=12cm \epsfbox{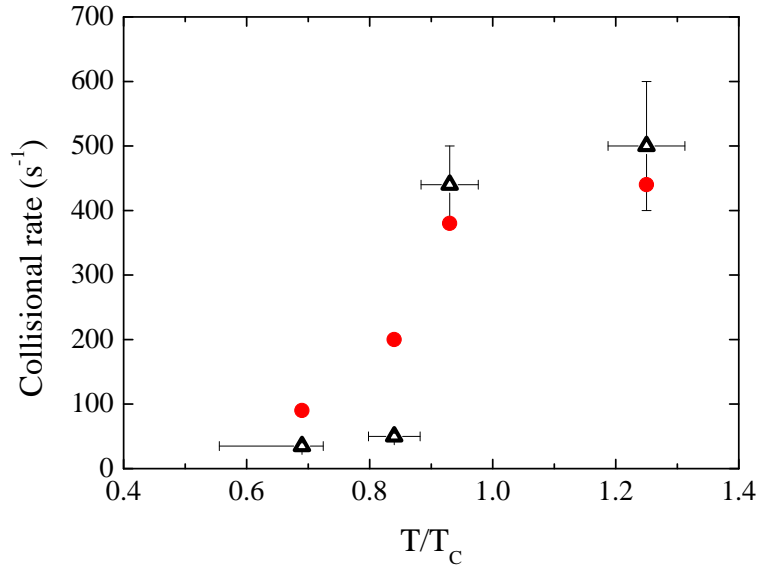}
\end{center}
\caption{\label{riass} Decrease of the fermion-boson collisional
rate as a function of the temperature along a typical evaporation
to BEC.  For $T<~T_c$, the experimental rates (triangles) are
smaller than the classical prediction (circles) due to the Pauli
blocking and superfluidity of K and Rb atoms respectively.}
\end{figure}

The variation of the collisional rate as a function of the reduced
boson temperature is shown in Fig.~\ref{riass}, for different
mixed samples in a typical evaporation across the phase-transition
to BEC. The number of Rb atoms varies from about 2$\times 10^5$ at
the highest temperature to 10$^4$ at the lowest temperature
because of losses due to evaporation, whereas the number of K
atoms stays almost constant ($N_K \approx 2\times 10^{4}$,
corresponding to a Fermi temperature $T_F \approx 2 T_c$). To
understand the role of the varying temperature and atom number on
the measured collisional rate, we have computed for each
experimental point the corresponding expected rate for a classical
gas (circles in Fig.~\ref{riass}). This collisional rate is
calculated from Eq.~\ref{gamma}, assuming that both K and Rb atoms
follow a Boltzmann distribution at all temperatures. The classical
prediction is clearly larger than the observed $\Gamma_{coll}$ for
$T<T_c$, indicating that quantum degeneracy of the mixed system is
influencing the fermion-boson interaction.

Apart from the Pauli blocking already addressed above, one should
also consider the superfluid behavior of the BEC \cite{ketterle}.
Collisions between the BEC and fermions are indeed suppressed if
the momentum exchanged in the collision is below $M_{Rb}c$, where
$c$ is the sound velocity
$c$=$\sqrt{4\pi\hbar^2a_{Rb}n_0/M_{Rb}^2}$. Note in particular the
dramatic reduction of $\Gamma_{coll}$ when lowering the
temperature from $T_c$ to 0.85 $T_c$, which cannot be reproduced
by the classical prediction.  Since the change in temperature
between these two measurements is negligible for fermions, the
reduction in collisional rate cannot be ascribed to Pauli blocking
in the Fermi gas, and this appears to be the consequence of
superfluidity in the BEC.

To gain a more quantitative insight into this phenomenon we study
in detail the measurement at $T$=0.85 $T_c$ in which the condensed
and uncondensed fractions are detectable. To model the system in
these conditions, we modify Eq.~\ref{model} to include a third
harmonic oscillator, i.e. the uncondensed Rb cloud, which is
coupled to both  the BEC and the Fermi gas. In the experiment we
were able to follow the center-of-mass motion of both the
condensed and uncondensed components of the Rb sample. The two
motions occurred at the same frequency and exactly in phase,
although they were out of phase with the K motion. Including this
observation, we can therefore write a simplified model that takes
into account only the relevant coupling terms
\begin{eqnarray}
\ddot{x}_{Rb} =& -\omega_{Rb}^2 x_{Rb}- \frac{4}{3} \frac{M_K}{M}
\frac{N_K}{N} \Gamma_{coll} \left(
\dot{x}_{Rb} - \dot{x}_K \right)  \\
\ddot{x}_K =& -\omega_K^2 x_K + \frac{4}{3} \frac{M_{Rb}}{M}
\left[\frac{N_{Rb}}{N} \Gamma_{coll} + \frac{N_{th}}{N_{th}+N_K}
\Gamma_{th} \right]\left( \dot{x}_{Rb} - \dot{x}_K \right)\,,
\nonumber \label{modelb}
\end{eqnarray}
where $N$ and $M$ are the total number of atoms and mass of the
Fermi gas and condensed Rb fraction, and $N_{th}$ is the
uncondensed Rb fraction. Here the thermal Rb cloud moves together
with the BEC, and interacts with the Fermi gas through the
additional damping rate $\Gamma_{th}$. Since the measurement was
performed in almost exactly same conditions as the measurement of
Fig. \ref{fermitherm} (i.e. at the same temperature and with the
same number of atoms in the Fermi gas and in the uncondensed Rb
cloud), we can fix $\Gamma_{th}$ to the nominal value determined
in that experiment, $\Gamma_{coll}$=60~s$^{-1}$. Using this value
we can unambiguously determine the collisional rate between atoms
in the Fermi gas and the condensed fraction of the BEC,
$\Gamma_{coll}$=35(10)~s$^{-1}$. This value is consistent with the
one measured directly in the sample where the uncondensed fraction
is not detectable. A damping mechanism that would be taken into
account in this condition is the Landau damping. This has been
introduced to describe the damping of the relative motion of the
BEC condensed and uncondensed components \cite{pita, ketterle,
ferlaino}. It could be interesting to extend the study of this
mechanism to the Fermi-Bose case. Although the observed damping
rates already give evidence of a non-classical behaviour of
Fermi-Bose dipolar collisions in the degenerate regime, further
modelling is needed to extract quantitative information about the
system. One phenomenon that must also be taken into account is the
breaking of superfluidity \cite{raman}, since the relative
velocity between the BEC and the Fermi gas is comparable to the
sound velocity of our system $c\approx $ 2 $\mu$m/ms during
collisions.

 \section{Discussion and conclusions}

In conclusion we have investigated for the first time coupled
dipolar oscillations in a Fermi-Bose dilute gas composed of K and
Rb atoms. The study of the damping of the oscillations in thermal
samples was demonstrated to be a powerful tool for the accurate
determination of the interspecies scattering length. The large
value of the scattering length in this K-Rb pair allowed to access
both the collisionless and hydrodynamic collisional regimes by
varying the temperature and density of the samples.

At the onset of quantum degeneracy we observe a reduction of the
collisional rate. In the interaction of a thermal bosonic gas with
the degenerate Fermi gas only the Pauli blocking of collisions is
expected to play a role.  In the presence of BEC, superfluidity
has the effect of lowering the collisional rate. Further modelling
of a Fermi-Bose mixture could allow one to extract from the
experimental observations useful information on the mean-field
interactions \cite{yabu,sogo} and on the temperature of the
mixture.

\section{Acknowledgements}
We acknowledge useful discussions with A. Simoni. This work was
supported by MIUR, by EC under the Contract HPRICT1999-00111, and
by INFM, PRA "Photonmatter" and by the European Science Foundation
through the "BEC2000+" programm.

\end{document}